\documentclass{iau}
\usepackage{graphicx}
\title[Proceedings IAU Symposium No. 294, 2013] 
{Venus transit, aureole and solar diameter}
\author[Wenbin Xie, Costantino Sigismondi, Xiaofan Wang \& Paolo Tanga]   
{Wenbin Xie$^1$$^,$$^a$, Costantino Sigismondi$^2$, Xiaofan Wang$^1$$^,$$^b$\\
 \and Paolo Tanga$^3$}
\affiliation{$^1$ Key Laboratory of Solar Activity, National Astronomical Observatories, \\
Chinese Academy of Sciences, email: {\tt $^a$wb\_xie@bao.ac.cn} {\tt $^b$wxf@nao.cas.cn} \\[\affilskip]
$^2$ ICRA/Sapienza Universit{\`a} di Roma, Ateneo Pontificio Regina Apostolorum,
\\email: {\tt sigismondi@icra.it}
$^3$ Observatoire de la C\^{o}te d'Azur, email: {\tt paolo.tanga@oca.eu}
}
\pubyear{2013}
\volume{294}  
\jname{Solar and Astrophysical Dynamos and Magnetic Activity}
\editors{A.G. Kosovichev, E.M. de Gouveia Dal Pino \& Y. Yan, eds.}
\begin{document}
\maketitle
\begin{abstract}

The possibility to measure the solar diameter using the transits of Mercury has been exploited to investigate the past three centuries of its evolution and to calibrate these measurements made with satellites.
This measurement basically consists to compare the ephemerides of the internal contact timings with the observed timings.
The transits of Venus of 2004 and 2012 gave the possibility to apply this method, involving a planet with atmosphere, with the refraction of solar light through it creating a luminous arc all around the disk of the planet.
The observations of the 2012 transit made to measure the solar diameter participate to the project Venus Twilight Experiment to study the aureole appearing around it near the ingress/egress phases.

\end{abstract}

\section{Introduction}

Shapiro in 1980 and Parkinson et al., 1980 calculated the solar diameter after the historical data on Mercury transits. In Fig. 1 are reported these measurements (triangles) along with two squares representing the last two determination of the solar diameter made with SOHO satellite with the transits of 2003 and 2006 by Emilio et al., 2012.
The debate around the variations of the solar diameter is still open and also we have presented some measurements made with SDO/HMI satellite during 2011, in order to show the reliability of these measurements (Wang and Sigismondi, 2013): in the process of calculation of the solar diameter as seen from 1AU there are still some residual oscillations of the same timescale of the orbital periods of the satellite around the Earth. This may be done also to other instrumental effects, than the orbital ones.

\section{The Venus Twilight Experiment}

During the Venus transit in 2004 several observers collected data useful to the characterization of the mesosphere of the planet, by observing the solar light refracted at the corresponding altitude range. The "aureole" thus formed, is observable during the ingress and egress phases of the transit, when Venus is crossing the solar limb. For the 2012 opportunity we prepared a set of coronagraphs to obtain multi-wavelength, space- and time-resolved photometry of the aureole, in collaboration with other space- and ground-based campaigns. The coronagraphs were distributed in the visibility area around the Pacific, over eight sites where local logistic support and scientific expertise were present.
Several sites obtained useful data at frame rates of several images/sec at 450, 535, 607 and 760 nm (FWHM 10 nm). A comparison with data collected at the 2004 transit shows that variations in the aspects of the aureole are present.
These can be linked to variations in the vertical distribution of the absorbers (aerosols and cloud-top level; Tanga, et al., 2012).
The consequences of the aureole on the evaluations of the contact timings observed either from the ground either from space is expected to be of the order of the point-spread function of the telescope.
The aureole is extended also to the side of the planet immersed in the photosphere background and it decreases the perceived diameter of Venus,
modifying slightly the contact timings of the transit.

\begin{figure} \begin{center}
 \includegraphics[width=5.2in]{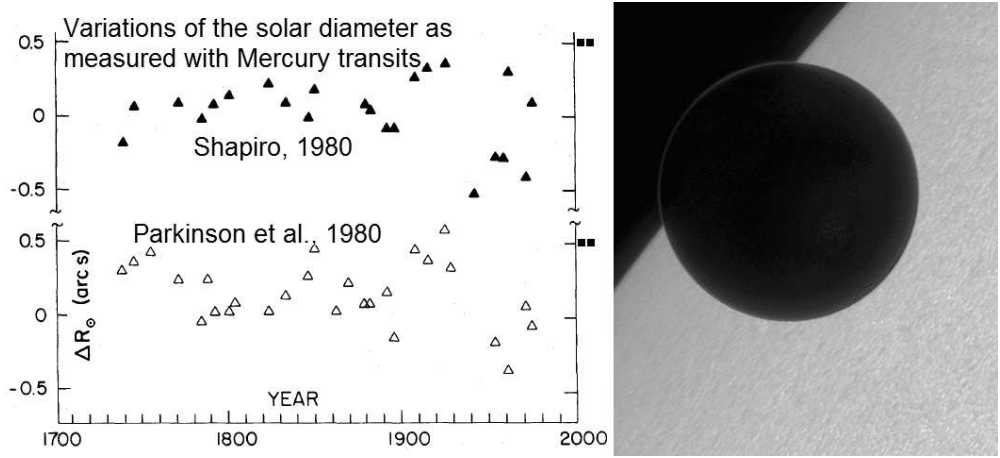}
 \caption{The values of the solar diameter obtained from the visual data of the transits of Mercury (filled and empty triangles) according to two different data analysis (Shapiro 1980 and Parkinson, et al. 1980). The last two data are the transits of 2003 and 2006 observed with SOHO and reduced by Emilio et al. 2012. On the right side there is an image of the transit of Venus 2012 from HINODE satellite, with the aureole.
The scatter of the measurements may be due both to an intrinsic variation of the solar diameter and to the difference between the instruments adopted for the observations, as it was in the case of 1832.
Bessel (1832) observed the transit of Mercury of 5 May 1832 with a 16-cm Fraunhofer heliometer and perceived the diameter of Mercury as 6.7 arcsec and
Gambart (1832) observed with a 6.7-cm Dollond refracting telescope, measuring the diameter of the planet as 9.3 arcsec; from the ephemerides the diameter of Mercury was 6.9 arcsec, and both the observations were done under optimal weather conditions. The Point-Spread Function of the telescope ruled a major role on the determination of the contact timings, more affecting the observations made with smaller instruments (Sigismondi, 2011).
}
   \label{fig1}
\end{center}
\end{figure}

\section{Conclusions}

The large amount of data obtained at the Huairou Solar Observing Station and from space satellite available for the 2012 transit is still under analysis and we expect to have soon a reliable value for the solar diameter on that date. Unfortunaltely we have not the same abundance of data for the 2004 transit, because virtually no chronodated sequences of images of ingress and egress phases are available.

\end{document}